\documentclass[12pt]{elsarticle}
\usepackage[T1]{fontenc}
\usepackage[utf8]{inputenc}
\usepackage{mathptmx}
\usepackage{pdfpages}       
\usepackage[labelfont=bf]{caption}
\usepackage{geometry}                		
\usepackage{graphicx}				
\usepackage{amssymb}
\date{}

\date{}

\journal{Neuron}

\begin{document}




# Modular Segregation of Structural Brain Networks Supports the Development of Executive Function in Youth

Graham L. Baum,[a] Rastko Ciric,[a] David R. Roalf,[a] Richard F. Betzel,[c]
Tyler M. Moore,[a] Russell T. Shinohara,[b] Ari E. Kahn,[c] Megan Quarmley,[a]
Philip A. Cook,[d] Mark A. Elliott,[d] Kosha Ruparel,[a] Raquel E. Gur,[a] Ruben C. Gur,[a]
Danielle S. Bassett,[c,e]* & Theodore D. Satterthwaite [a]*

Departments of Psychiatry,[a] Biostatistics and Epidemiology,[b] Bioengineering,[c] Radiology,[d] and Electrical and Systems Engineering[e] of the University of Pennsylvania, Philadelphia PA 19104 USA

*=Co-senior authors

Address correspondence to:
Theodore D. Satterthwaite, M.D., M.A.
10th floor, Gates Building
Hospital of the University of Pennsylvania
34th and Spruce Street
Philadelphia, PA 19104

**SUMMARY:** The human brain is organized into large-scale functional modules that have been shown to evolve in childhood and adolescence. However, it remains unknown whether structural brain networks are similarly refined during development, potentially allowing for improvements in executive function. In a sample of 882 participants (ages 8-22) who underwent diffusion imaging as part of the Philadelphia Neurodevelopmental Cohort, we demonstrate that structural network modules become more segregated with age, with weaker connections between modules and stronger connections within modules. Evolving modular topology facilitated network integration, driven by age-related strengthening of hub edges that were present both within and between modules. Critically, both modular segregation and network integration were associated with enhanced executive performance, and mediated the improvement of executive functioning with age. Together, results delineate a process of structural network maturation that supports executive function in youth.

# INTRODUCTION

Modularity is a fundamental feature of complex systems, including social groups, cyber-physical systems, and diverse biological networks (Newman, 2006). A network module is a group of densely interconnected nodes, which often are the basis for specialized subunits of information processing (Sporns and Betzel, 2016). Functional neuroimaging studies have demonstrated that the human brain has a well-defined modular organization, as reflected in the presence of large-scale functional networks (Biswal et al., 1995; Damoiseaux et al., 2006; Power et al., 2011; Yeo et al., 2011). While the exact number and spatial distribution of functional network modules varies somewhat by analytic approach, a remarkable convergence exists across independent datasets and laboratories (Damoiseaux et al., 2006; Power et al., 2011; Yeo et al., 2011).

Commonly described modules include somatomotor (Biswal et al., 1995), visual (Corbetta et al., 1998), default mode (Raichle et al., 2001; Buckner et al., 2008), and fronto-parietal systems (Dosenbach et al., 2007; Vincent et al., 2008). While brain network modules emerge very early in life (Fair et al., 2008; van den Heuvel et al., 2015; Thomason et al., 2013), a growing body of work has shown that these functional modules are refined during youth. During childhood and adolescence, functional modules become more distinct: connectivity within modules increases while connectivity between modules declines (Power et al., 2010; Fair et al., 2007; Fair et al., 2008; Fair et al., 2009; Dosenbach et al., 2010; Gu et al., 2015; Satterthwaite et al., 2013b; Supekar et al., 2009; Anderson et al., 2011). Such development allows for functional specialization, reducing interference among systems (Fornito et al., 2012) and facilitating cognitive performance (Hampson et al., 2010). Modularity is particularly relevant for executive function, which relies on co-activation of executive regions and reciprocal suppression of non-executive regions such as the default mode network (Anticevic et al., 2010; Barber et al., 2013; Persson et al., 2007; Satterthwaite et al., 2013a). Thus, available data suggests that development of network modularity may serve as a substrate for the evolution of executive capability during youth.

Despite convergent evidence for the developmental emergence of functional network modularity, there is relatively scant data regarding the maturation of underlying structural brain networks that support this functional architecture (Hermundstad et al., 2014). Prior work demonstrates substantial correspondence between functional and structural measures of brain connectivity (Goñi et al., 2014; Honey et al., 2009; Mišić et al., 2016), although structural connections tend to be a subset of densely connected, polysynaptic functional networks (Betzel et al., 2014; Hermundstad et al., 2013). Structural networks in adults are highly modular (Bassett et al., 2010; Bassett et al., 2011), but it remains unknown if this topology evolves substantially during youth. Correspondence between functional and structural data intuitively suggests that, like functional networks, structural networks should become increasingly segregated during development. However, prior studies using relatively small samples report conflicting results, including declining modularity (Chen et al., 2013), increasing modularity (Chen and Deem, 2015; Huang et al., 2015), or no change with age (Hagmann et al., 2010b; Lim et al., 2015). Larger sample sizes may be necessary for resolving the variability of findings reported in previous studies.

Beyond this mixed data regarding normative developmental trends, the impact of structural network development on cognitive performance remains poorly described. Cognitive capability improves substantially during youth, with executive function undergoing a protracted phase of development throughout adolescence and young adulthood (Gur et al., 2012; Luna et al., 2004). Describing how structural brain networks evolve to support executive function is necessary to understand the basis for many sources of adolescent morbidity and mortality, which are prominently associated with failures of executive function (Shamosh et al., 2008; Casey et al., 2008; Casey et al., 2011). Finally, such data are a prerequisite for studies of neuropsychiatric disorders, which are increasingly understood as disorders of brain development (Insel, 2010; Rapoport et al., 2012; Casey et al., 2014), are marked by executive dysfunction (Shanmugan et al., 2016), and are linked to the disruption of evolving network topology (Alexander-Bloch et al., 2012; Di Martino et al., 2014; Castellanos and Proal, 2012; Voineskos et al., 2010).

Here we sought to define the normative development of structural network modules, and delineate the impact of modular maturation on executive functioning. We tested the hypothesis that modules within structural brain networks become more segregated with age, as seen in functional brain networks. Further, we predicted that segregated structural modules would support enhanced executive functioning. To address these hypotheses, we capitalized upon a large sample of 882 youths who completed diffusion imaging as part of the Philadelphia Neurodevelopmental Cohort (PNC), a community-based study of brain development that includes rich neuroimaging and cognitive data (Satterthwaite et al., 2014a; Calkins et al., 2015; Satterthwaite et al., 2015). As described below, results provide novel evidence that structural brain networks undergo a process of modular segregation analogous to prior accounts of functional network development. Critically, these data reveal that the refinement of structural network modules mediate the development of executive function.

**RESULTS**

We investigated the evolution of structural brain networks in a sample of 882 youth aged 8-22 who completed neuroimaging as part of the PNC (**Figure 1A**). As expected, executive function improved markedly with age (**Figure 1B**). Structural brain networks were constructed using nodes defined based on a parcellation of each subject's structural image into 234 anatomically defined regions; structural connectivity between these nodes was estimated using deterministic tractography (**Figure 2**). Each network node was assigned *a priori* to one of the functional network modules defined by Yeo et al. (2011). Although these module partitions were defined in an independent dataset, using a different imaging modality, the modularity quality of the functional partition imposed on subject-level structural connectivity matrices ($Q_{Yeo}$) was highly significant ($p<1\times10^{-10}$). Furthermore, data-driven analysis of structural networks using community detection procedures produced network modules that showed significant similarity to the *a priori* functional modules ($p<1\times10^{-10}$; **Figure S1**).

*Segregation of structural network modules increases with age*

We first sought to understand whether structural network modules became more segregated with age. To do this, we calculated the average participation coefficient for each subject's network. The participation coefficient quantifies the relative balance of a brain region's

between-module versus within-module connectivity (Guimerà and Nunes Amaral, 2005). *Greater* modular segregation is therefore indicated by *lower* participation coefficient values, with reduced between-module and elevated within-module connectivity. We examined the development of modular segregation using a generalized additive model with penalized splines, which allows for statistically rigorous modeling of both linear and non-linear effects while minimizing over-fitting (Wood, 2004; Wood, 2011). Based on emerging evidence that diffusion-weighted imaging measures are systematically biased by motion artifact (Yendiki et al., 2013), we included in-scanner motion as a covariate in all models in addition to participant sex. To ensure that results reflected changes in network topology, rather than global differences in network connectivity, total network strength was also included as a covariate in all analyses (Li et al., 2012).

The participation coefficient declined significantly with age (**Figure 3A**; $p<1\times10^{-10}$), indicating enhanced modular segregation. Developmental increases in modular segregation were differentially distributed across modules (**Figure 3B**), with the most robust declines observed in the somatomotor and default mode modules. To further understand which regions were driving these effects, we examined the participation coefficient of individual network nodes. As expected, the nodal participation coefficient declined in many regions (**Figure 3C**), with many of the most significant reductions occurring in regions within the default mode system. Two exceptions to this overall trend were observed, with increasing participation coefficient in the right rostral frontal gyrus and frontal operculum.

Next, we investigated the degree to which developmental effects on modular segregation were driven by changes in within-module connectivity, between-module connectivity, or both. We found that both effects were significant: within-module connectivity increased with age (**Figure 4A**; $p<1\times10^{-10}$), whereas between-module connectivity declined (**Figure 4B**; $p<1\times10^{-10}$). Moreover, modular segregation was reflected in individual network edges (**Figure 4C**), with permutation-based analysis revealing that a higher proportion of connections that strengthened with age were located within a module (**Figure 4D**; $p<0.001$).

*Results are robust to methodological approach*

Given this strong evidence for developmental modular segregation, we next pursued a set of analyses to determine if our results were dependent upon specific methodological choices. We evaluated multiple parameters, including alternative measures of network segregation, higher-resolution node systems, and different edge measures. First, we examined a complementary measure of modular segregation: the modularity quality index of the Yeo et al. partition ($Q_{Yeo}$) applied to individual subject-level structural connectivity matrices. For this metric, greater modular segregation is denoted by a greater $Q_{Yeo}$ value. As expected, this analysis revealed that modular segregation increased with age (**Figure 5A**; $p=1.06\times10^{-9}$). Second, while our primary results used the functional partition defined by Yeo et al. (2011), we additionally evaluated modular segregation using a data-driven partition of the structural connectome (see **Figures S1 and S2**). As seen in **Figure 5B**, age-related decline in the mean participation coefficient remained evident ($p<1\times10^{-10}$). Third, we directly calculated the modularity quality index of a data-driven partition of each subject's structural connectivity matrix ($Q_{subj}$). We observed that this statistic – which provides an individualized measure of modularity that is not dependent on a group-level partition – also increased significantly with age (**Figure 5C**; $p=0.0007$), indicating greater modular segregation. Fourth, we investigated the impact of using a more fine-grained

4

network parcellation (n=463 nodes). This did not impact the observed results, with age-related declines in the participation coefficient remaining highly significant (**Figure 5D**; $p<1\times10^{-10}$). Fifth, we evaluated developmental effects when using different measures of structural connectivity instead of mean fractional anisotropy. Results using raw streamline count (**Figure 5E**; $p=6.52\times10^{-7}$) or volume-normalized streamline density (**Figure 5F**; $p=4.56\times10^{-8}$) remained highly similar. Lastly, in order to rule out the possibility that observed results were driven by potentially confounding variables, we included additional model covariates such as total brain volume, handedness, race, and maternal education; results were unchanged ($p<1\times10^{-10}$). Conversely, results were consistent when covariates such as total network strength, sex, and motion were removed from the model ($p<1\times10^{-10}$).

*Modular segregation contributes to global network efficiency*

Having established that network modules become more segregated with age, and that this finding was not dependent on specific analytic choices, we evaluated the impact of evolving network modularity on measures of network integration. Global network efficiency ($E_{glob}$) provides a measure of network integration by quantifying information flow across a network as the shortest path between pairs of nodes (Bassett et al., 2009). In many networks, modularity and global efficiency are inversely related, as network segregation by module partitions extends the path length. However, in some cases it is possible for networks to become both more modular *and* more efficient; this unusual situation occurs when connectivity within modules is efficiently organized and hub edges form strong links between otherwise segregated modules (Sporns and Betzel, 2016). To determine which scenario characterized human neurodevelopment, we first examined the relationship between global efficiency and age while controlling for the covariates described previously. Replicating previous reports (Chen et al., 2013; Hagmann et al., 2010b), we found that global efficiency increases with age (**Figure 6A**; $p<1\times10^{-10}$). Next, we calculated the correlation between modular segregation (mean participation coefficient) and global efficiency, while co-varying for age to control for shared developmental trends. Mean participation coefficient was negatively associated with global efficiency (**Figure 6B**; $p<1\times10^{-10}$), suggesting that the development of network modules does not result in fragmentation, but rather is associated with global network integration.

*Age effects are concentrated in hub edges that promote network integration*

To better understand this highly specialized association between network modularity and efficiency, we evaluated the edge betweenness centrality for each network connection. Edge betweenness identifies hub connections by providing a measure of how much a given network edge lies upon the shortest path of communication through a network, and thus contributes to global efficiency. Here we defined *hub edges* as those connections within the top quartile of edge betweenness across all network edges. Critically, edges that strengthened with age were enriched for hub edges ($p<0.001$; see **Figure 6C**). Both within- ($p<0.001$) and between-module ($p<0.001$) edges that strengthened with age had higher betweenness than expected by chance (**Figure 6D**; see **Supplemental Information)**. Furthermore, the average strength of all within-module (**Figure 6E**; $p<1\times10^{-10}$) and between-module (**Figure 6F**; $p<1\times10^{-10}$) edges that strengthen with age was associated with global efficiency, suggesting that developmental effects are concentrated within connections that facilitate network integration. The striking combination of increasing modular segregation and enhanced global efficiency demonstrates that structural brain networks

5

become more modular *and* more integrated in development. These dual processes are driven by selective strengthening of network hub edges, which are present within network modules and also provide critical links between increasingly segregated modules.

*Modular segregation mediates development of executive function in youth*

Next, we evaluated the cognitive implications of modular segregation by examining associations with individual differences in executive function. Mean whole-brain participation coefficient was associated with improved executive performance ($p=0.018$). At the level of individual modules, we found that segregation of the frontoparietal control system was uniquely associated with executive ability (**Figure 7A**; $p=0.005$), suggesting a network-specific substrate for executive function. As a final step, we examined whether age-related changes in executive function and modularity were related. Mediation analyses revealed that this was indeed the case (**Figure 7B**; $p=0.006$), suggesting that the development of segregated structural brain modules mediates the age-related improvement in executive function. These mediating effects were specifically driven by the frontoparietal module ($p=0.012$). Similarly, global efficiency was associated with executive functioning ($p=0.037$), and also mediated executive development ($p=0.002$).

To evaluate the specificity of these results, we examined associations with other domains of cognition, such as social cognition and memory performance. While no association with memory was found, modular segregation was also significantly associated with social cognition ($p=0.022$), which was driven by segregation of the default mode module ($p=0.012$). This effect mediated improvements in social cognition with age ($p=0.008$). Together, these results demonstrate that developmental segregation of specific structural network modules may support the development of disparate cognitive domains.

**DISCUSSION**

Capitalizing on a large sample of youth imaged as part of the PNC, we demonstrated that modules within human structural brain networks become increasingly segregated with age. This result was robust to specific methodological choices, and driven by a combination of enhanced within-module connectivity and declining between-module connectivity. Age related changes were concentrated within specific hub edges, allowing for networks to simultaneously become more modular and more integrated with age. Critically, segregation of network modules mediated the development of executive function during adolescence.

*Segregation of structural network modules parallels development of functional networks*

The delineation of robust, reproducible large-scale functional networks has had a tremendous impact on human neuroscience research (Power et al., 2011; Yeo et al., 2011). As a result, functional network modules have evolved to become the dominant framework by which human imaging data is interpreted. The conceptualization of the brain as a modular entity has had a particularly pronounced effect on theories of development, where convergent results have shown that functional network modules are present early in life (Thomason et al., 2013; van den

Heuvel et al., 2015), and continue to develop during youth (Power et al., 2010; Fair et al., 2007; Fair et al., 2008; Fair et al., 2009; Dosenbach et al., 2010; Gu et al., 2015; Satterthwaite et al., 2013b; Supekar et al., 2009; Anderson et al., 2011). In contrast, smaller studies of structural brain networks have produced heterogeneous results regarding the development of structural network modules that have not aligned well with functional imaging data (Chen et al., 2013; Chen and Deem, 2015; Hagmann et al., 2010b; Huang et al., 2015; Lim et al., 2015). When considered in light of prior studies that have reported substantial correspondence between brain structure and function (Goñi et al., 2014; Honey et al., 2009; Mišić et al., 2016), the disparity between developmental accounts of structural and functional network modules has been difficult to reconcile.

Leveraging a large sample imaged as part of the PNC, we were able to resolve this discrepancy by demonstrating that structural network modules develop in a similar manner as functional brain networks, and become increasingly segregated with age. Modular segregation was present at every scale evaluated, including the whole network, individual network modules, and specific network nodes. Importantly, results were robust to a variety of analytic choices regarding network nodes, edges, and modules; such methodological replication is critical as parameter choices may sometimes impact inference (Hagmann et al., 2010a). For example, while we employed a commonly used set of functional network modules which were defined *a priori,* analysis of data-driven structural modules provided highly convergent results.

Follow-up analyses revealed further parallels with functional imaging studies, and demonstrated that the process of modular segregation is driven by a combination of enhanced connectivity within a module as well as diminished connectivity between modules. Notably, the network modules that demonstrated the greatest developmental segregation were the somatomotor and default mode modules. Prior work has shown that both the default mode and somatomotor systems are highly segregated systems, with a low participation coefficient (Power et al., 2011). The relative isolation of these specific systems in the network may reflect a high degree of processing specialization (Power et al., 2013). The present results thus suggest that the differential evolution of structural network modules is similarly driven by each module's network role.

*Modular networks become more integrated through strengthening of hub edges*

In many networks, modular segregation is associated with reduced network integration, as measured by global efficiency. We found that this was not the case in development, and that increasing modularity was in fact associated with enhanced network integration. This robust association was the result of targeted strengthening of specific hub edges. These hub edges were present within but also between modules, allowing for integration across increasingly segregated partitions. These results are congruent with prior studies that have demonstrated that connections between network hubs strengthen preferentially with age (Baker et al., 2015), and that network efficiency increases during development (Chen et al., 2013; Hagmann et al., 2010b). The present data emphasize that increasing modular segregation does not result in isolation of functional sub-systems, but is associated with global network integration through strengthening of hub edges that facilitate both intra- and inter-module connectivity.

7

*Structural network maturation supports the development of executive function in youth*

Having defined a normative process of modular segregation, we evaluated the cognitive impact of this developmental effect. While controlling for age, we found that greater modular segregation of structural networks was associated with better executive performance. Critically, modular segregation mediated the observed improvement of executive performance with age, and was driven by segregation of the frontoparietal module. Associations between module segregation and cognition were domain-specific: segregation of the default mode mediated age-related improvements in social cognition, which is reliant on regions within that network (Buckner and Carroll, 2007; Schacter et al., 2007). The process of structural network segregation may allow for functional specialization, and reduce competitive interference between brain systems (Fornito et al., 2012). Such a process is suggested by convergent data from task-based fMRI, which has shown that individual differences in performance are related to selective recruitment of executive regions and suppression of activity elsewhere (Anticevic et al., 2010; Barber et al., 2013; Persson et al., 2007; Satterthwaite et al., 2013a). Furthermore, building on prior work that reported an association between intelligence and the global efficiency of structural (Li et al., 2009) and functional networks (van den Heuvel et al., 2009) in relatively small adult samples, we found that global efficiency also mediated developmental improvements in executive function. Taken together, the current data suggest that structural brain networks re-configure with age, becoming both more modular and more integrated. This specific topology may allow for both functional specialization within modules as well as coordination across modules, which is necessary for effective implementation of dynamic executive processes (Hutchison and Morton, 2015; Braun et al., 2015).

*Limitations*

Notwithstanding the strengths of this study, several limitations should be noted. First and foremost, this is a cross-sectional dataset, which has inherent limitations for studies of development (Kraemer et al., 2000). The mediating role that network maturation plays in the development of executive function could be further interrogated using longitudinal data. These limitations offer clear directions for additional research. Ongoing follow-up of this cohort will yield informative data, as will other large-scale studies of brain development, including the IMAGEN consortium (Schumann et al., 2010), the NKI-Rockland sample (Nooner et al., 2012), and the forthcoming Adolescent Brain and Cognitive Development Study. Finally, it should be noted that diffusion-based tractography remains limited in its ability to fully resolve the complex architecture of the structural connectome (Jbabdi et al., 2015).

*Conclusions*

In this report, we demonstrated that structural brain modules become more segregated with age. Strengthening of specific within- and between-module hub edges allowed for a simultaneous process of network integration that evolves in concert with modular segregation. Finally, both modular segregation and global network integration mediated the development of executive function in youth. These data resolve an ongoing debate in the field regarding the normative development of structural brain networks, and delineate an important new mechanism for the development of executive functioning in youth. These findings may be relevant for understanding how individual differences in brain development associate with risk-taking behaviors, which are linked to failures of executive function, and are a major source of morbidity

and mortality in adolescence (Shamosh et al., 2008; Casey et al., 2008; Casey et al., 2011). Furthermore, as both abnormalities within developing networks and executive system dysfunction (Shanmugan et al., 2016) are a common feature of diverse types of psychopathology (Alexander-Bloch et al., 2012; Di Martino et al., 2014; Castellanos and Proal, 2012; Voineskos et al., 2010), structural network development may evolve to become an important imaging biomarker of risk and resilience during the critical period of adolescence.

# EXPERIMENTAL PROCEDURES

*Participants*

This study included 882 subjects between 8 and 22 years of age (mean age=15.06, SD=3.15; 389 males, 493 females) who were imaged as part of the PNC (Satterthwaite et al., 2014a; Satterthwaite et al., 2015). Participants were excluded from analyses due to gross structural brain abnormalities (Gur et al., 2013), a history of inpatient psychiatric treatment, current use of psychotropic medications, and medical disorders that could impact brain function. Participants were only included if both structural (Vandekar et al., 2015) and diffusion images (Roalf et al., 2016) passed rigorous quality assurance procedures. All participants completed the Penn computerized neurocognitive battery, which included 14 tests (Gur et al., 2002; Gur et al., 2012). Cognitive performance was summarized by a recent factor analysis (Moore et al., 2014) of both speed and accuracy data, which delineated three factors corresponding to the efficiency of executive function, episodic memory, and social cognition. For further details see **Supplemental Information**.

*Image acquisition & processing*

All high-resolution structural and 64-direction diffusion images were collected on the same 3T Siemens TIM Trio scanner using the same sequences (Satterthwaite et al., 2014a). The T1 image was processed using FreeSurfer version 5.3 (Fischl, 2012), and parcellated into 234 cortical and subcortical regions (Cammoun et al., 2012). The diffusion images were distortion corrected, skull stripped, and motion and eddy current corrected with FSL's `eddy` tool (Jenkinson et al., 2012). Native-space T1 parcels were dilated by 4mm to extend regions into white matter, and co-registered to the first non-weighted (b=0) volume using a boundary-based rigid body transformation (Greve and Fischl, 2009). The diffusion tensor was estimated in DSI Studio and whole-brain deterministic fiber tracking was implemented with 1,000,000 streamlines per subject after removing all streamlines with length less than 10mm (Yeh et al., 2013). Edge weights in the adjacency matrix were defined by mean fractional anisotropy along streamlines connecting each pair of nodes (Mišić et al., 2016; Bohlken et al., 2016; Baker et al., 2015). See **Figure 2** and **Supplemental Information** for more detail.

*Measurement of modular segregation*

Nodes were assigned to modules according to their overlap with seven functional networks defined *a priori* (Yeo et al., 2011); subcortical nodes were assigned to their own additional module. In addition to such *a priori* functional assignment, network modules were defined directly from the structural connectivity data using a generalized version of the Louvain community detection algorithm (Bassett et al., 2013; Blondel et al., 2008; Mucha et al., 2010). The similarity of data-driven partitions of the structural data were compared to the functional partition using the *z*-score of the Rand coefficient (Traud et al., 2011). Modular segregation was quantified using the participation coefficient, which measures the balance of between- versus within-module connectivity for each brain region (Guimerà and Nunes Amaral, 2005; Rubinov and Sporns, 2010). A higher participation coefficient at a given node indicates more between-module and less within-module connectivity. Each node's participation coefficient was averaged over modules and the whole brain in order to evaluate modular segregation at multiple scales. To

10

understand whether changes in within- or between-module connectivity were driving observed effects, we also calculated these measures separately for the whole brain and each network module (Gu et al., 2015). Also see **Supplemental Information**.

*Statistical analyses of age-related changes in modular segregation*

Linear and nonlinear effects of age were flexibly modeled with penalized thin-plate splines using generalized additive models (GAMs; Wood, 2004; Wood, 2011), which provide statistically rigorous analysis of non-linear effects while minimizing over-fitting. As in previous work (Satterthwaite et al., 2014b; Vandekar et al., 2015), nonlinearity was penalized within the GAMs using restricted maximum likelihood. Developmental models of each network measure were assessed by modeling age with a spline term, while including participant sex and in-scanner motion as covariates. To ensure that changes in global connectivity strength did not drive analyses of network topology, total network strength was also included as a covariate in all analyses (Li et al., 2012). Throughout, multiple comparisons were controlled using the False Discovery Rate ($q<0.05$; Genovese et al., 2002). Permutation testing was used to assess whether the *a priori* functional network partition fit subject-level structural connectivity matrices and whether within-module connections were enriched for age effects (see **Supplemental Information**).

*Methodological replications*

To verify that observed age-related increases in modular segregation were not simply due to specific processing choices, we repeated analyses of global network segregation using a variety of other parameters. Analyses were repeated using a subject-specific measure of modularity quality ($Q$) for both *a priori* functional networks and a data-driven structural partition. Using the functional network partition, we also examined age-related changes in modular segregation in a higher resolution node system (n=463 instead of n=234), and using two alternative edge definitions (streamline count and normalized streamline density). Finally, we evaluated the effect of additional model covariates (race, maternal education, handedness, and total brain volume). For additional details, see **Supplemental Information**.

*Relationship between modular segregation and global network efficiency*

Global efficiency was calculated for each participant's structural network (Latora and Marchiori, 2001; Rubinov and Sporns, 2010) and age-related effects were examined as above. The relationship between global efficiency and modular segregation was examined while controlling for age (and other covariates as above). To further describe how age-related changes in within- and between-module connectivity might drive measures of global efficiency, we calculated the edge betweenness centrality (Brandes, 2001; Rubinov and Sporns, 2010). Edge betweenness quantifies the degree to which each edge participates in the shortest paths within a network, and thus contributes to global efficiency. Normalized edge betweenness was calculated for each edge, split by type (within- versus between-module) and age effect (strengthens with age versus no change). Hub edges were defined as connections within the top quartile of edge betweenness across all network edges. Permutation tests were used to evaluate whether connections that strengthened with age were enriched for hub edges, and had higher edge betweenness than expected by chance (see **Supplemental Information**).

*Associations with executive function*

To determine if network segregation was related to executive performance, a GAM was used which incorporated the factor score for executive efficiency as well as model covariates (spline age, sex, in-scanner motion, total network strength). Specificity analyses additionally evaluated relationships with other cognitive domains including episodic memory and social cognition. Global efficiency was also assessed for relationships with cognition. Linear mediation analyses investigated whether age-related improvements in cognition were mediated by modular segregation or integration; significance was assessed using bootstrapping procedures (Preacher and Hayes, 2008).

**SUPPLEMENTAL INFORMATION**

Supplemental Information includes Figure S1, Figure S2, and Supplemental Experimental Procedures.

**ACKNOWLDGEMENTS**

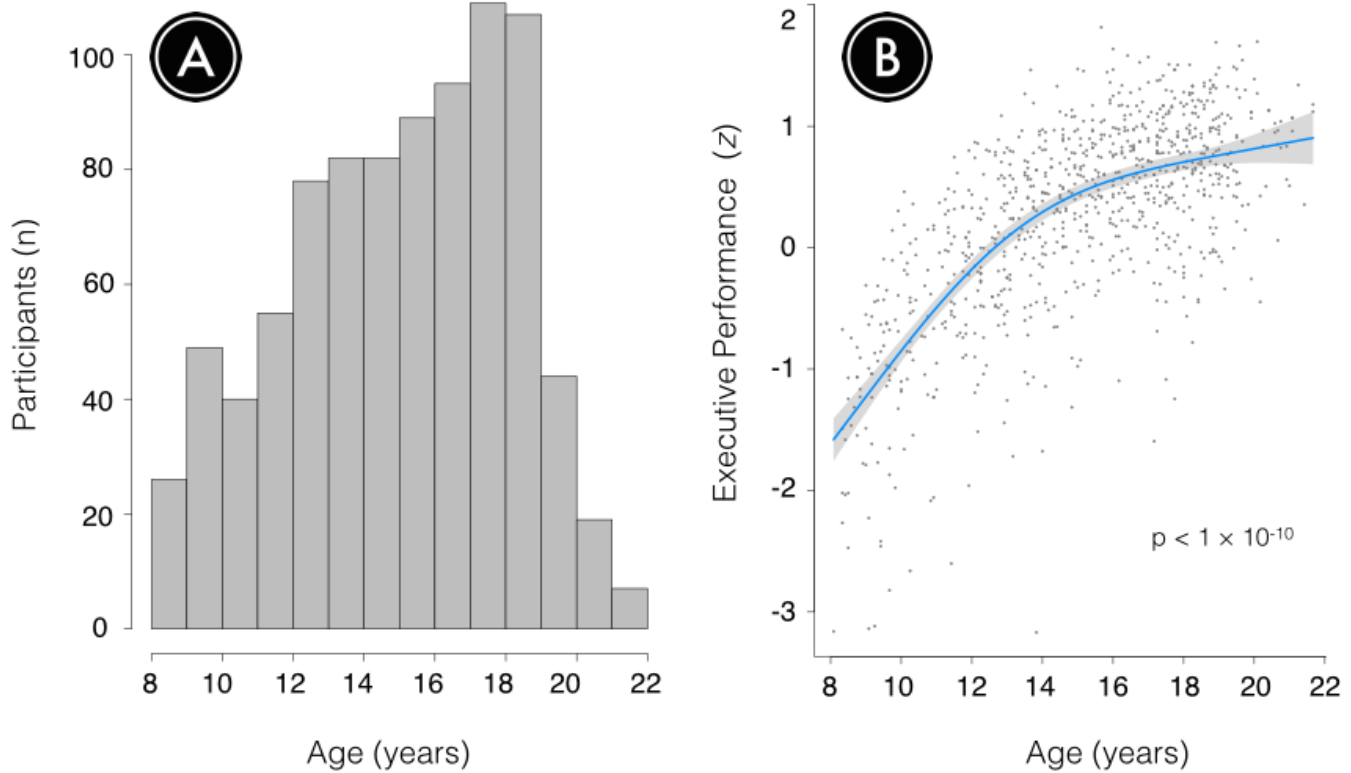

**Figure 1**. Executive functioning improves with age. **A.** Age distribution of 882 youth completing diffusion imaging as part of the PNC. **B**. Executive performance on a neurocognitive battery improves with age. Blue line represents the best fit from a general additive model; shaded area indicates 95% confidence interval. Model includes participant sex as a covariate.

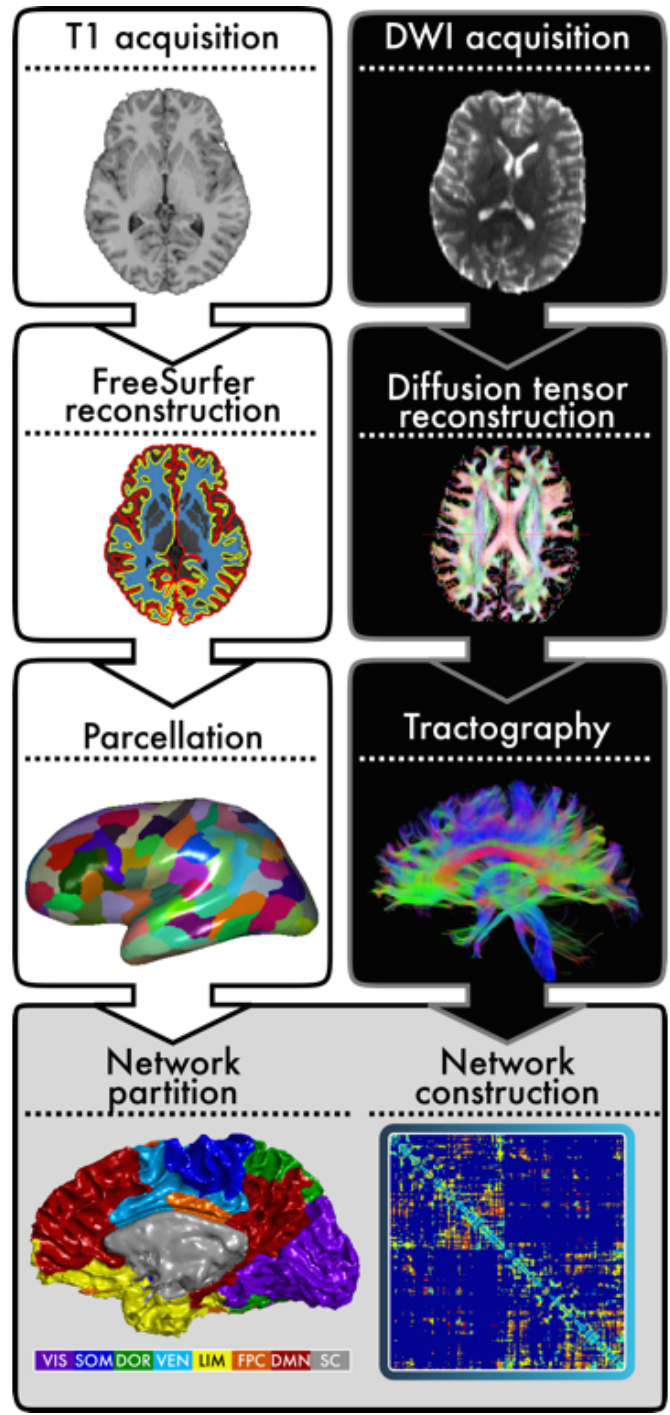

**Figure 2.** Connectome construction. For each subject, the T1 image was processed using FreeSurfer and parcellated into 234 network nodes on an individualized basis. Deterministic whole-brain fiber tracking was used to create a symmetric adjacency matrix (234×234), where the edge weight was defined as the mean fractional anisotropy (FA) along the connecting streamlines. Network nodes were each assigned to one of the seven large-scale functional modules defined by Yeo et al. (2011); subcortical nodes were assigned to an eighth modules. VIS=visual, SOM=somatomotor, DOR=dorsal attention, VEN=ventral attention, LIM=limbic, FPC=frontoparietal control, DMN= default mode network, SUB=subcortical.

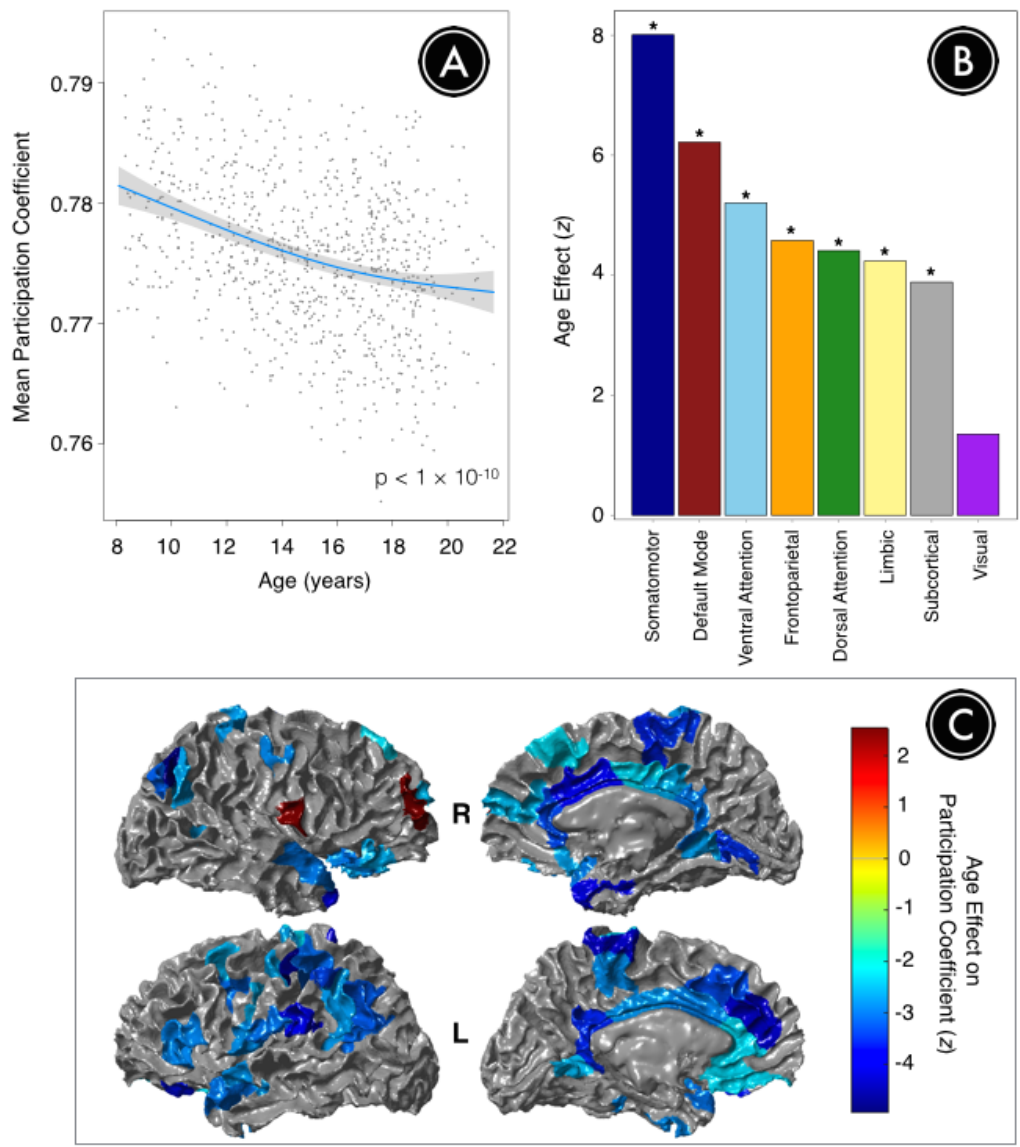

**Figure 3**. Structural brain network modules become increasingly segregated with age. **A.** Modular segregation was quantified as the mean participation coefficient across all network nodes, with lower values indicating more segregation. Participation coefficient values declined significantly with age. **B.** Modular segregation is differentially distributed across functional systems. Age-related modular segregation is most robust in the somatomotor and default mode systems, but also present in other networks. **C.** Age-related changes in participation coefficient provide convergent results for individual nodes, and demonstrate widespread declines with age. The strongest age-related reductions of the participation coefficient were seen in default mode regions such as the posterior cingulate. Two exceptions to this overall trend were the right rostral frontal gyrus and frontal operculum, where participation coefficient increased with age. Blue line represents the best fit from a general additive model; shaded area indicates 95% confidence interval. All analyses control for sex, in-scanner motion, and network strength.
Color palette represents $z$-transformed $p$ values from a general additive model. Images are thresholded to control for multiple comparisons using the False Discovery Rate ($Q<0.05$).
*indicates $p<0.001$.

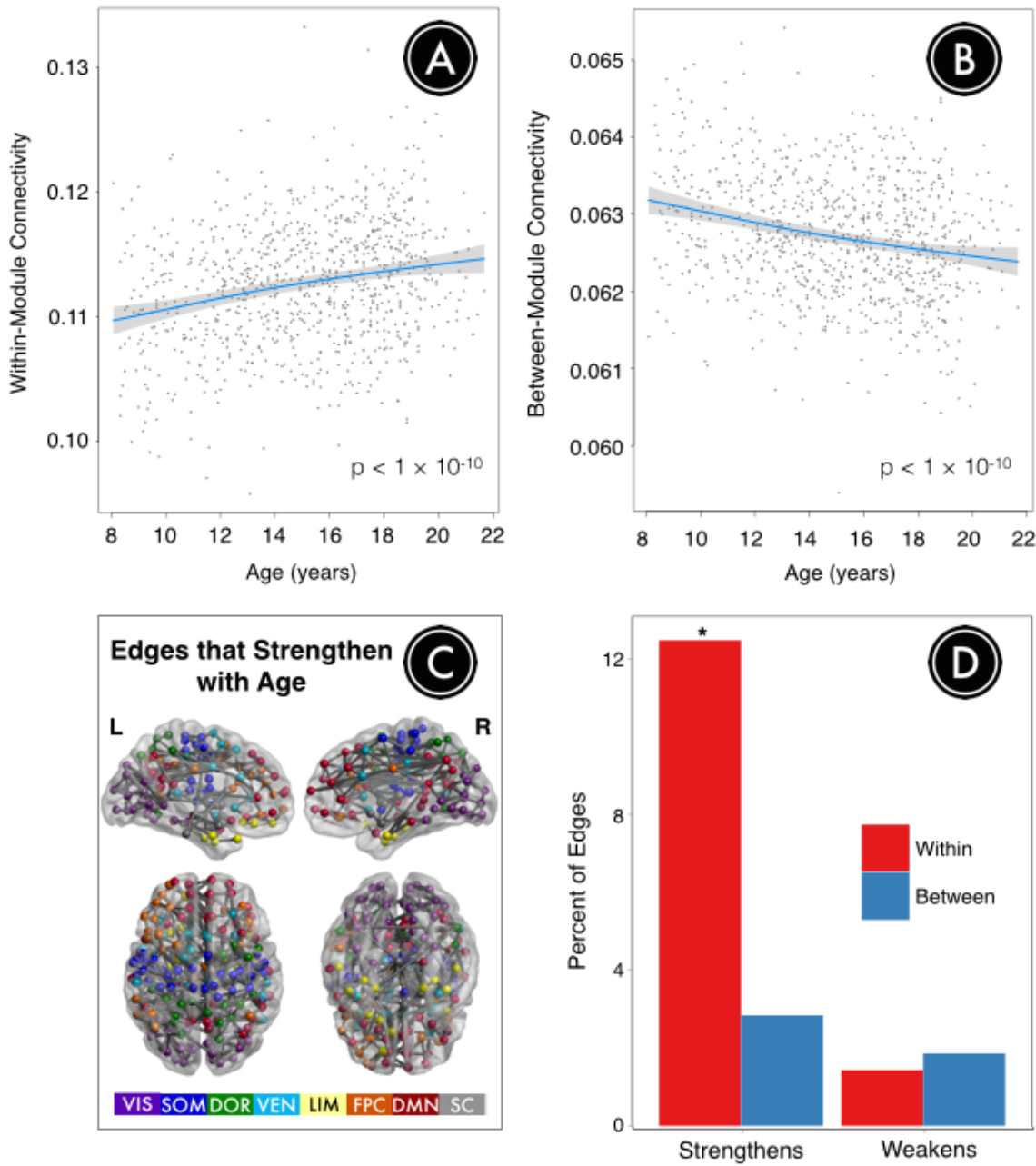

**Figure 4**. Modular segregation is driven by a combination of both enhanced within-module connectivity and reduced between-module connectivity. **A.** Average strength of within-module connectivity increases with age. **B.** Between-module connectivity decreases across development. **C.** Convergent effects are seen at the level of individual graph edges (image thresholded using Bonferroni corrected $p<0.05$ for clarity). **D.** A higher percentage of within-module connections (red) strengthen with age than expected by chance. All analyses include sex, in-scanner motion, and network strength as model covariates. * indicates $p<0.001$.

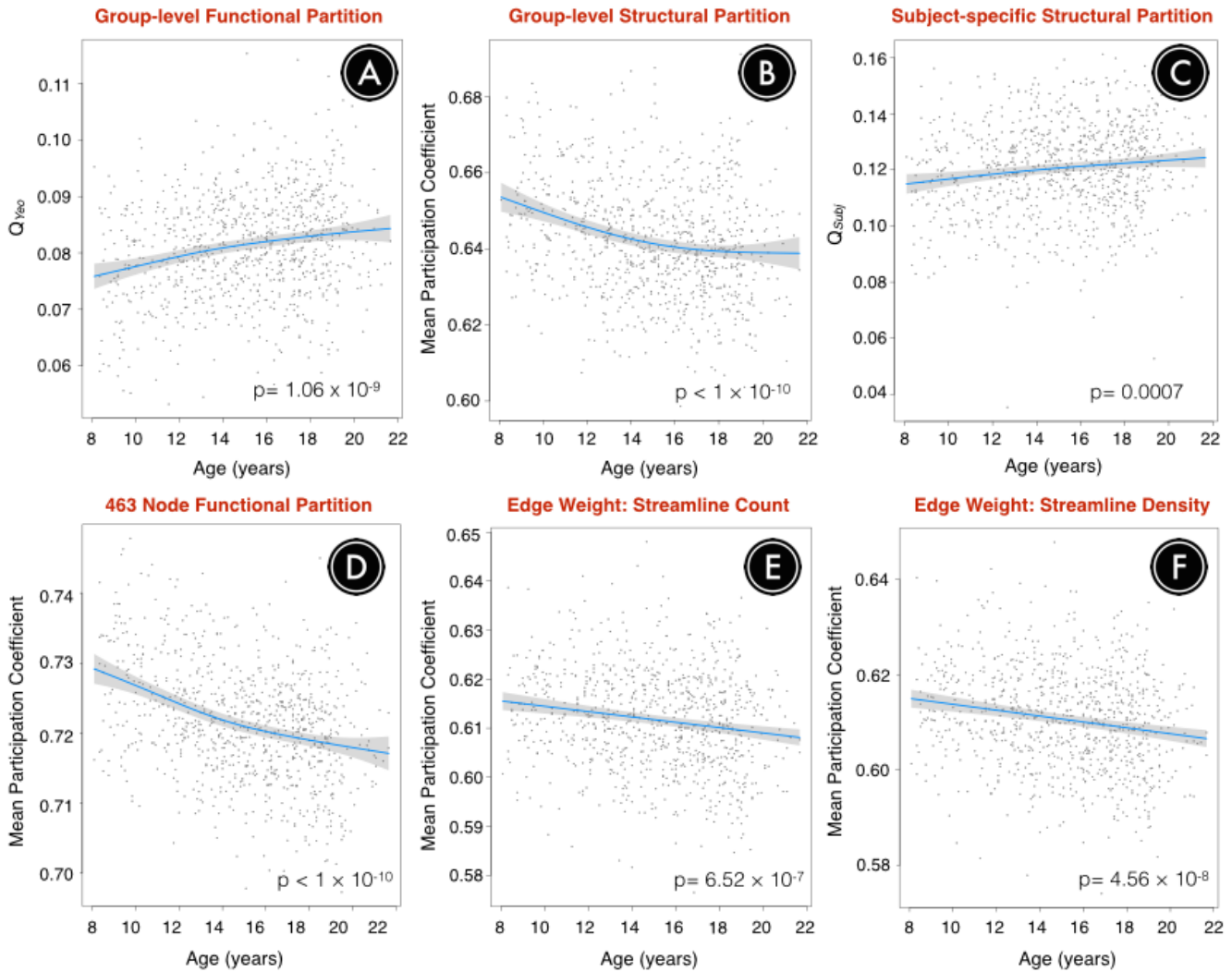

**Figure 5**. Results are robust to methodological choices. Regardless of specific processing decisions, an increase in modular segregation with age was observed. **A**. Convergent findings result when using an index of the modularity quality for the Yeo partition, where higher $Q$ indicates more segregated modules. **B**. When using a group-level structural partition, modular segregation (mean participation coefficient) decreases with age. **C**. Modularity quality of subject-level connectivity matrices also increases with age. **D**. Results remain unaffected when a higher-dimensional parcellation is used (n=463 nodes), **E**. when streamline count is used instead of FA as an edge weight, and **F**. when normalized streamline density is used as the edge weight. Lower participation coefficient indicates more segregated modules. Blue line represents the best fit from a general additive model; shaded area indicates 95% confidence interval. All models include sex, in-scanner motion, and total network strength as covariates.

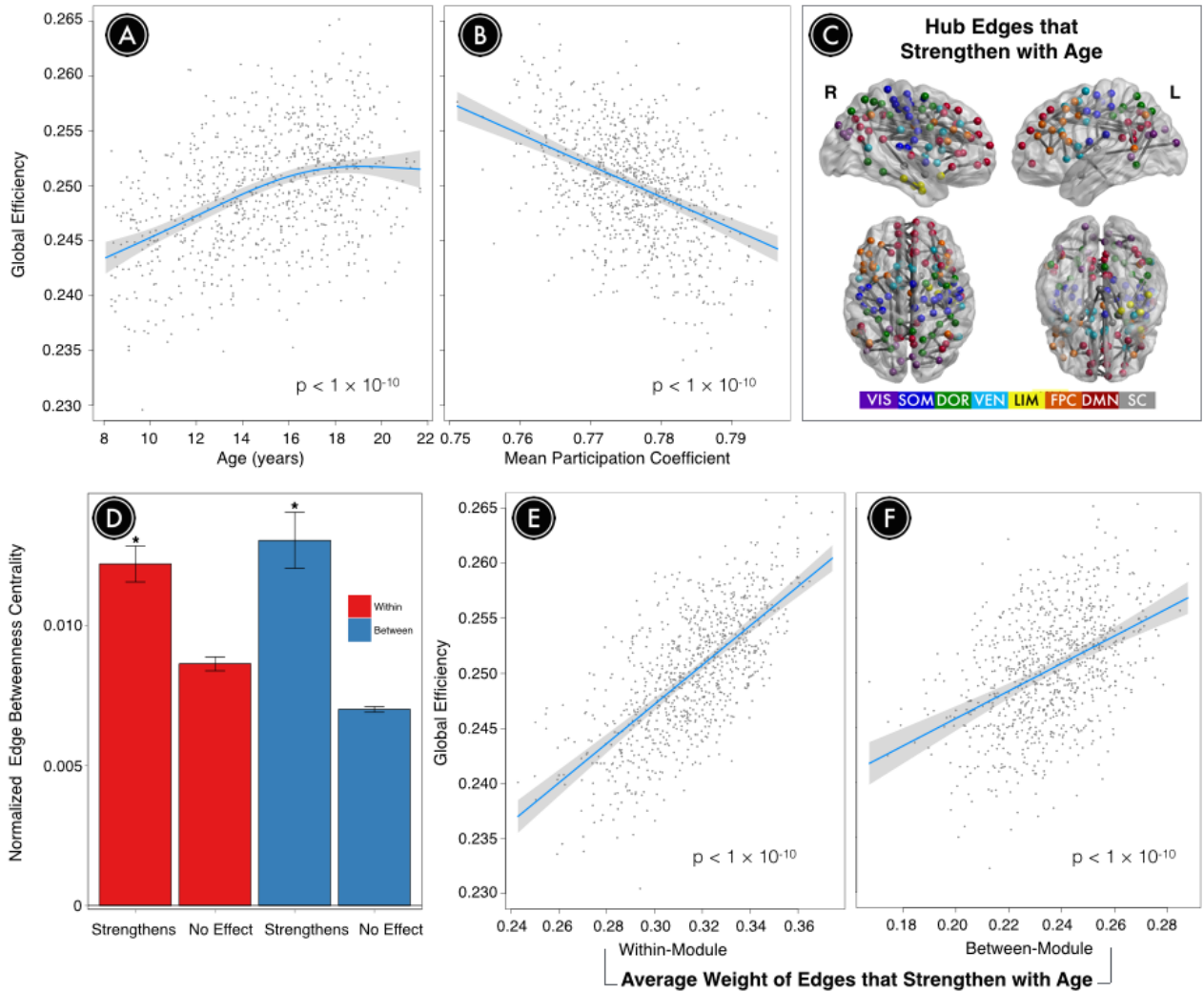

**Figure 6**. Modularity is associated with network integration, and is driven by developmental strengthening of specific hub edges. **A.** Replicating prior work, global network efficiency increases with age. Model includes sex, in-scanner motion, and total network strength as covariates. **B.** While controlling for age, lower mean participation coefficient is associated with greater network efficiency, indicating a positive association between modular segregation and network integration. **C.** Connections that strengthen with age are enriched for hub edges (47%). Hub edges are defined as connections in the top quartile of edge betweenness centrality, which quantifies how often a given edge lies on the shortest path between nodes and thus facilitates global efficiency. Image thresholded using Bonferroni corrected $p<0.05$ for clarity. **D.** Both within- and between-module connections that strengthen with age have higher edge betweenness centrality than expected by chance. The average weight of within- (**E**) and between-module edges (**F**) that strengthen with age are positively associated with global efficiency. Blue line represents the best fit from a general additive model, shaded area indicates 95% confidence interval; * indicates $p<0.001$. Error bar represents standard error of the mean.

25

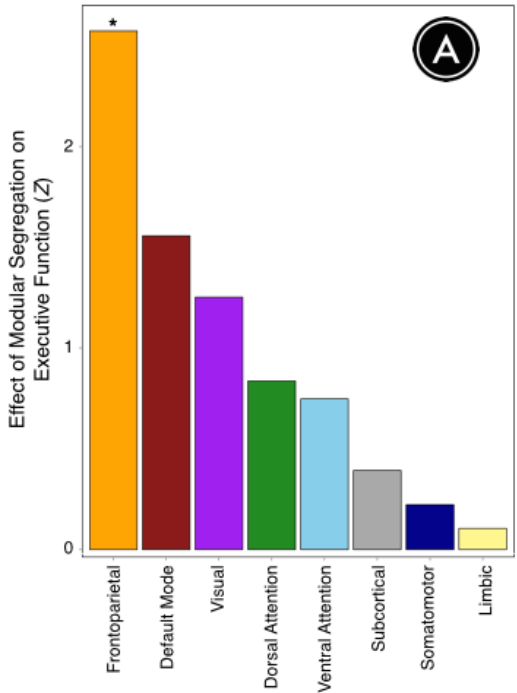
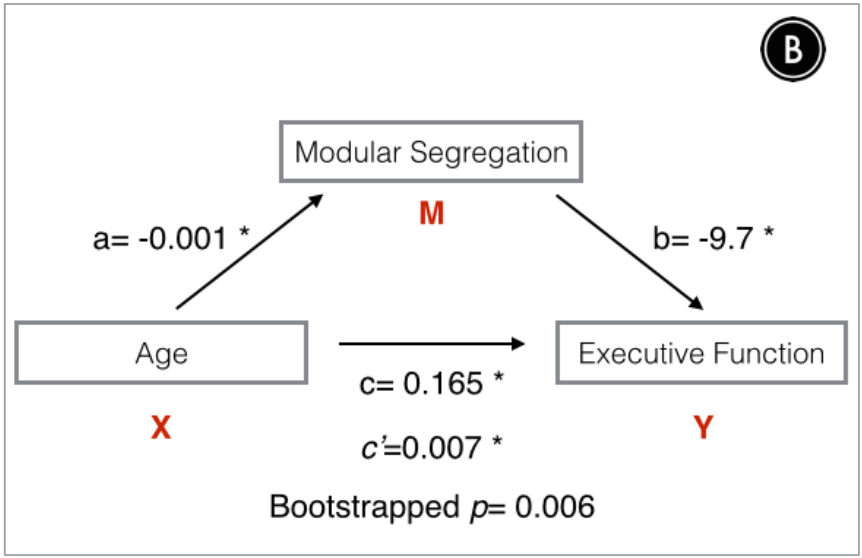

**Figure 7**. Segregation of structural modules supports the development of executive function in youth. **A.** While controlling for age, greater modular segregation in the frontoparietal control network is associated with better executive performance. **B.** Segregation of structural modules mediates the improvement of executive function with age. Mediation results shown as unstandardized regression coefficients. Significance of indirect effect ($c'$=0.007) was assessed using bootstrapped confidence intervals [0.002-0.012]. All models also include sex, in-scanner motion, and total network strength as covariates. * indicates $p<0.01$.

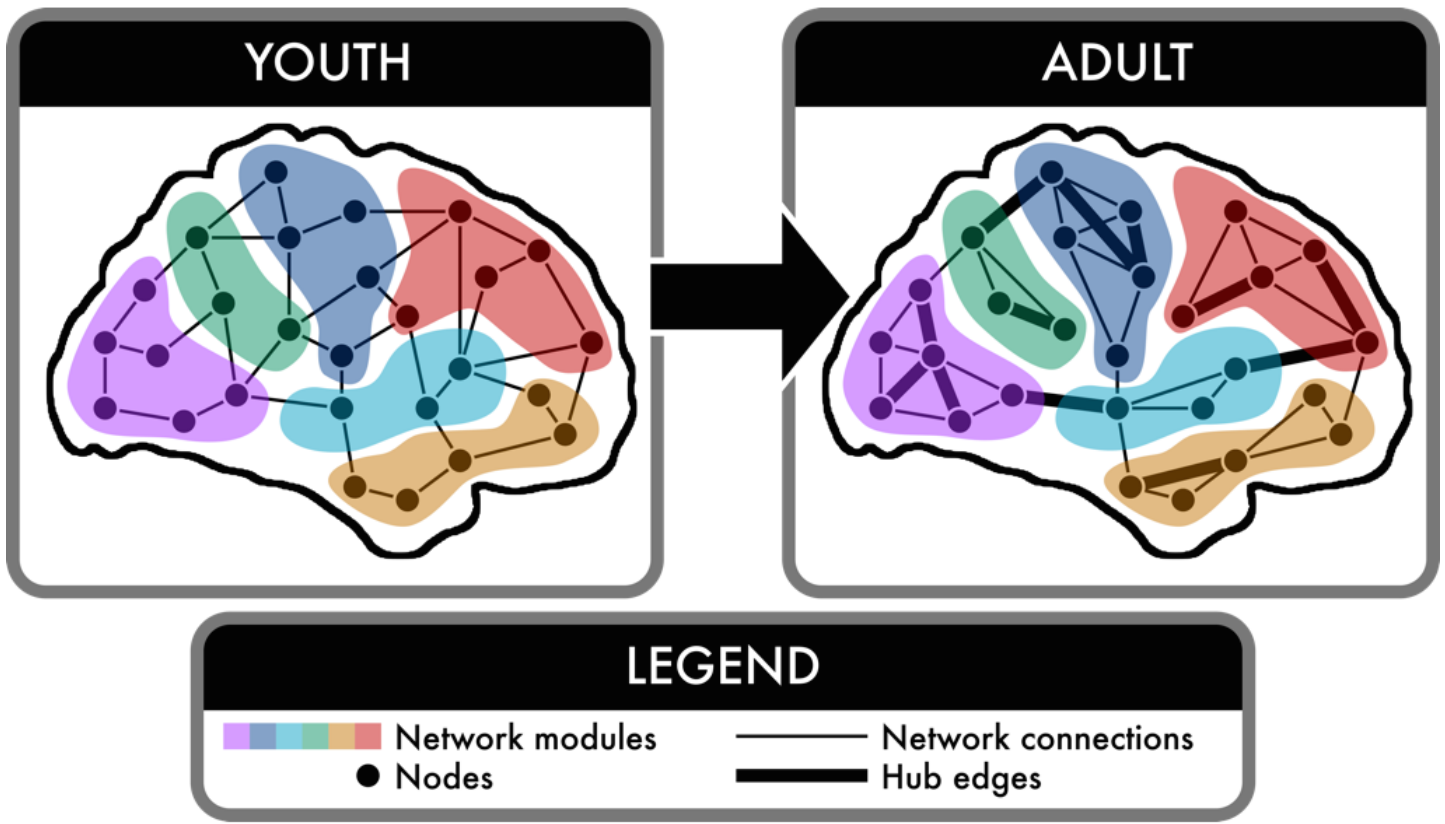

**Figure 8**. Modular evolution of structural brain networks across youth. From childhood through adulthood, structural brain networks become increasingly modular. The targeted strengthening of specific hub edges facilitates specialized information processing within distinct modules, and simultaneously enhances integration across modules. Hub edges are indicated by thick connections.

# SUPPLEMENTAL INFORMATION



\newpage
\section*{Supplemental Figures}
\begin{figure*}[h!]
\begin{center}
\includegraphics[width=\textwidth]{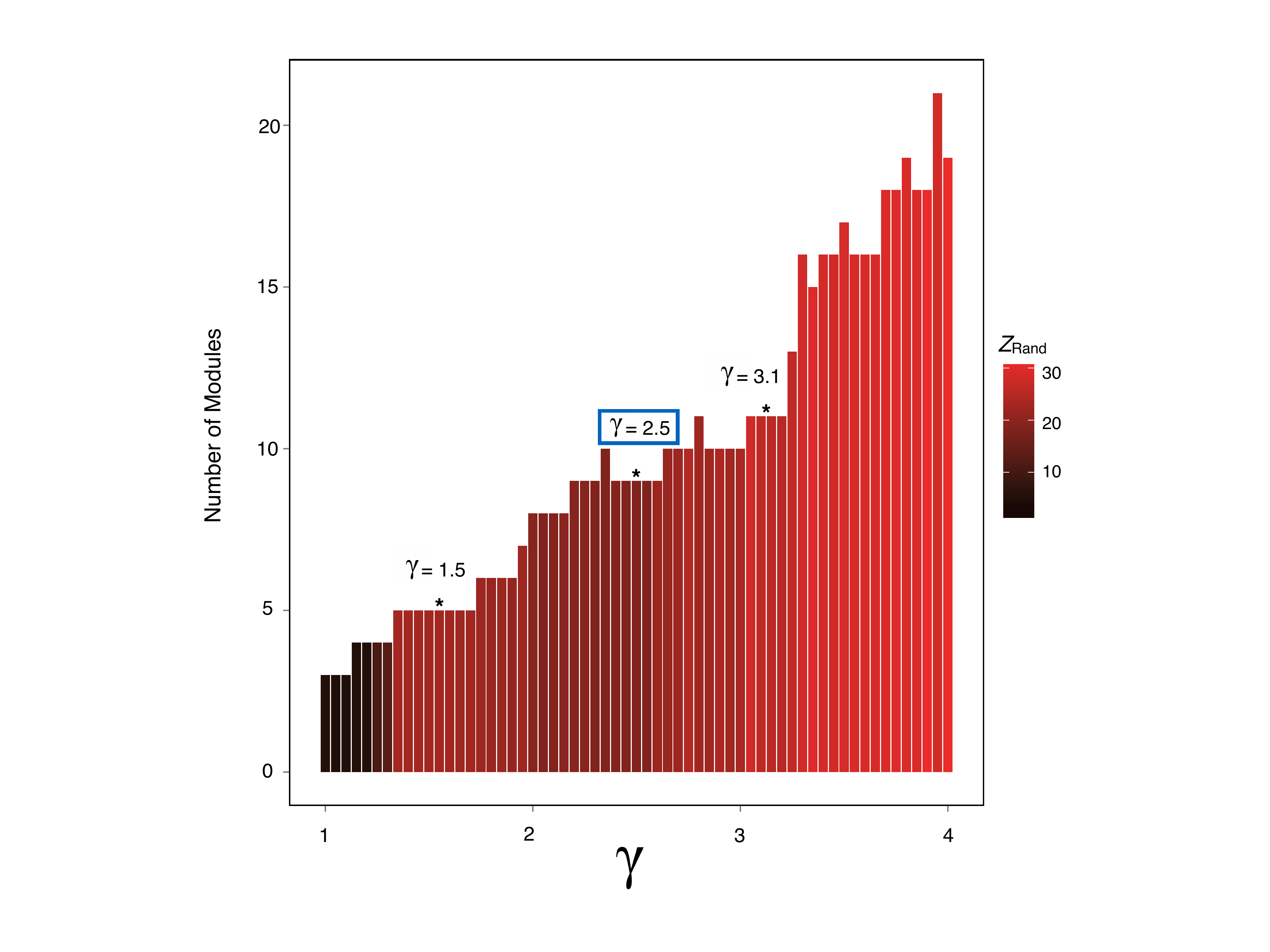}
\vspace*{-12mm}
\caption{\textbf{Number of modules identified in group-level structural partitions.} To examine alternative data-driven modular partitions of structural brain networks, we varied $\gamma$ over the interval $\left[0,4\right]$ in increments of 0.05. The number of modules identified in group-level consensus partitions increases as a function of $\gamma$. The similarity between structural partitions and \textit{a priori} functional partitions also increases with $\gamma$ and the number of identified structural modules. $\ast$ indicates alternative structural partitions identified at plateaus for the number of modules. Bars are colored by the \textit{z}-score of the Rand coefficient, which quantifies the similarity between structural partitions and the \textit{a priori} functional partition used throughout the main text. The 9-module structural partition identified at $\gamma$=2.5 (marked by blue box) is used to examine age-related effects on modular segregation in \textit{Figure 5}. The \textit{z}-score of the Rand coefficient is equal to 17.6 (p $<$ $1 \times 10^{-10}$) for this structural partition, suggesting a significant similarity with the functional partition beyond chance.}
\end{center}
\end{figure*}

\newpage
\begin{figure*}[h!]
\begin{center}
\includegraphics[width=\textwidth]{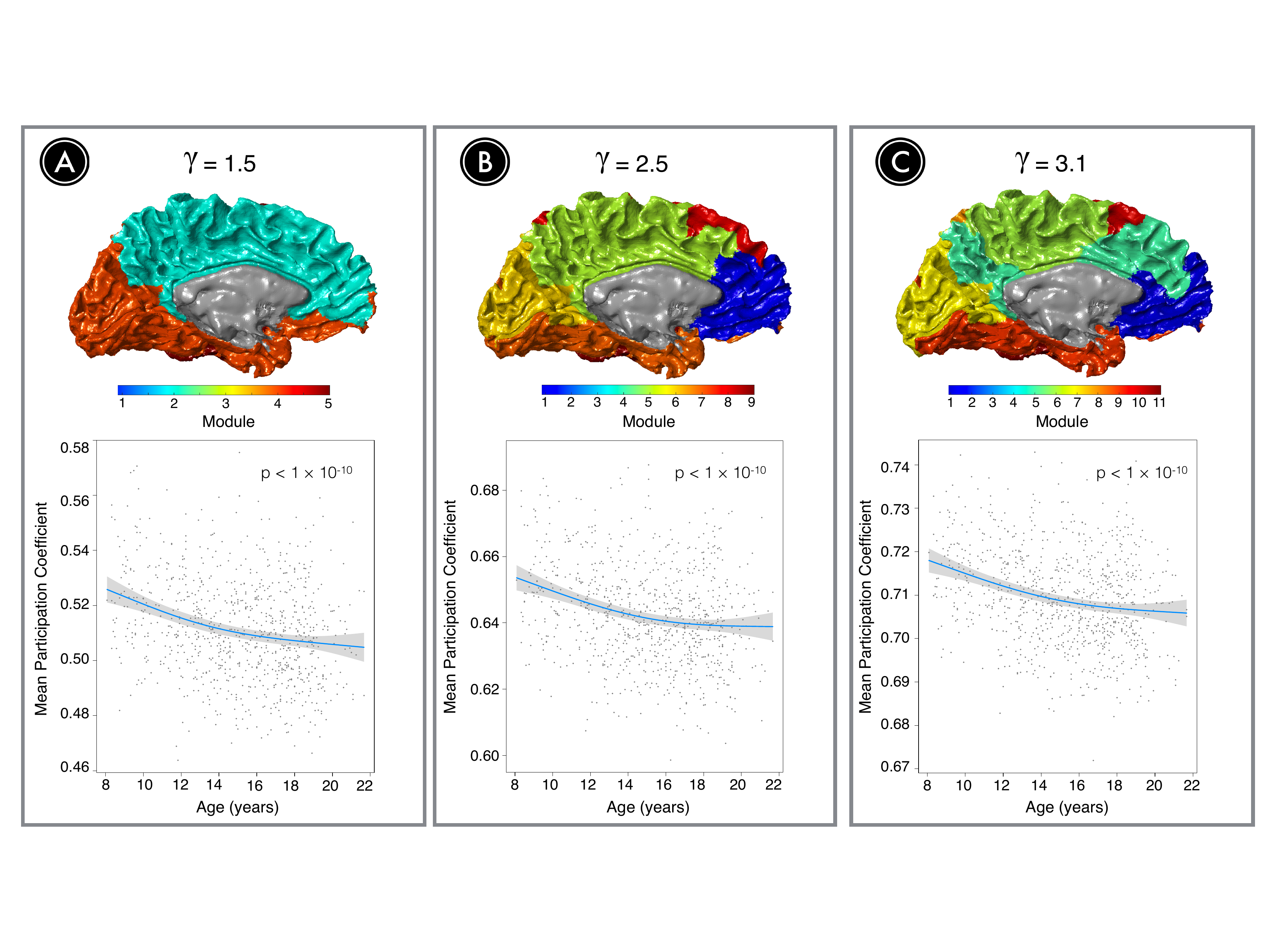}
\vspace*{-13mm}
\caption{ \textbf{Data-driven structural network modules become more segregated across youth.} Here we demonstrate that regardless of the group-level consensus partition used to define modules, modular segregation increases with age, as demonstrated by a significant decrease in the mean participation coefficient. This developmental pattern is replicated using a 5-module partition (\textbf{A}, $\gamma$=1.5), a 9-module partition (\textbf{B}, $\gamma$=2.5), and an 11-module partition (\textbf{C}, $\gamma$=3.1). The 9-module partition pictured in \textbf{B} is used to calculate modular segregation in \textit{Figure 5}. Blue line represents the best fit from a general additive model; shaded area indicates 95$\%$ confidence interval. Models include participant sex, in-scanner head motion, and total network strength as covariates.}
\end{center}
\end{figure*}

\newpage
\section*{Supplemental Experimental Procedures}

\subsection*{Subjects}
Diffusion tensor imaging (DTI) datasets were acquired as part of the Philadelphia Neurodevelopmental Cohort (PNC), a large community-based study of brain development. 1601 subjects completed the cross-sectional neuroimaging protocol (Satterthwaite et al., 2014). Datasets from 244 individuals were considered unusable due to incomplete acquisition or incidental findings. The remaining 1357 participants underwent a rigorous manual and automated quality assurance protocol for DTI datasets (Roalf et al., 2016), which flagged 157 subjects for poor data quality (e.g., low temporal signal-to-noise ratio). Of the remaining 1210 participants, 93 were flagged by automated quality assurance for low quality or incomplete FreeSurfer reconstruction of T1-weighted images. Of the remaining 1117 participants, 235 subjects were excluded for meeting any of the following criteria: gross radiological abnormalities, history of medical problems that might affect brain function, history of inpatient psychiatric hospitalization, use of psychotropic medication at the time of data acquisition, missing data, and/or high levels of in-scanner head motion (mean relative displacement between non-weighted volumes $>$ 2mm), which has been shown to impact measures derived from diffusion-weighted imaging (Roalf et al., 2016; Yendiki et al., 2013). These exclusions produced a final sample consisting of 882 youths (mean age=15.06, SD=3.15; 389 males, 493 females).

\subsection*{Cognitive Assessment}
The Penn computerized neurocognitive battery (Penn CNB) was administered to all participants. The CNB consists of 14 tests adapted from tasks applied in functional neuroimaging to evaluate a broad range of cognitive domains (Gur et al., 2002; Gur et al., 2012). These domains include executive control (abstraction and flexibility, attention, working memory), episodic memory (verbal, facial, spatial), complex cognition (verbal reasoning, nonverbal reasoning, spatial processing), social cognition (emotion identification, emotion intensity differentiation, age differentiation) and sensorimotor and motor speed. Accuracy and speed for each test were z-transformed. Cognitive performance was summarized by a recent factor analysis (Moore et al., 2014) of both speed and accuracy data, which delineated three factors corresponding to the efficiency of executive function, episodic memory, and social cognition.

\subsection*{Data Acquisition}
All MRI scans were acquired on the same 3T Siemens Tim Trio whole-body scanner and 32-channel head coil at the Hospital of the University of Pennsylvania. DTI scans were acquired using a twice- refocused spin-echo (TRSE) single-shot echo-planar imaging (EPI) sequence (TR = 8100ms, TE = 82ms, FOV = 240mm2 /240mm\textsuperscript{2} ; Matrix = RL: 128/AP:128/Slices:70, in-plane resolution (x and y) 1.875 mm$^2$; slice thickness = 2mm, gap = 0; flip angle = 90$^{\circ}$/180$^{\circ}$/180$^{\circ}$, volumes = 71, GRAPPA factor = 3, bandwidth = 2170 Hz/pixel, PE direction = AP). This sequence used a four-lobed diffusion encoding gradient scheme combined with a 90-180-180 spin-echo sequence designed to minimize eddy-current artifacts . DTI data were acquired in two consecutive series consisting of 32 diffusion encoding gradient schemes. The complete sequence consisted of 64 diffusion-weighted directions with b=1000s/mm$^2$ and 7 interspersed scans where \textit{b}=0 s/mm$^2$. The duration of DTI scans was approximately 11 minutes. The imaging volume was prescribed in axial orientation covering the entire cerebrum with the topmost slice just superior to the apex of the brain (Satterthwaite et al. 2014a). In addition to the DTI scan, a map of the main magnetic field (i.e., B0) was derived from a double-echo, gradient-recalled echo (GRE) sequence, allowing us to estimate field distortions in each dataset. 

\subsection*{Data Preprocessing}
Two consecutive 32-direction acquisitions were merged into a single 64-direction time-series. The skull was removed for each subject by registering a binary mask of a standard fractional anisotropy (FA) map (FMRIB58 FA) to each subject's DTI image using a rigid-body transformation (Smith et al., 2002). Eddy currents and subject motion were estimated and corrected using FSL's \textit{eddy} tool (Andersson and Sotiropoulos 2016). Diffusion gradient vectors were then rotated to adjust for subject motion estimated by \textit{eddy}. After the field map was estimated, distortion correction was applied to DTI data using FSL's FUGUE (Jenkinson et al., 2012). Lastly, DTI data was imported into DSI Studio software and the diffusion tensor was estimated at each voxel.

\subsection*{DTI Tractography}
Whole-brain fiber tracking was implemented for each subject in DSI Studio using a modified fiber assessment by continuous tracking (FACT) algorithm with Euler interpolation, initiating 1,000,000 streamlines after removing all streamlines with length less than 10mm or greater than 400mm (Yeh et al., 2013). Fiber tracking was performed with an angular threshold of 45$^{\circ}$, a step size of 0.9375mm, and a fractional anisotropy (FA) threshold determined empirically by Otzu's method, which optimizes the contrast between foreground and background (Yeh et. al., 2013). Diffusivity measures (e.g., FA, mean diffusivity, radial diffusivity, axial diffusivity) were calculated along the path of each reconstructed streamline. For each subject, tractography served as the basis for constructing structural brain networks.

\subsection*{Network Construction}
Following T1 reconstruction in FreeSurfer (version 5.3), cortical and subcortical gray matter was parcellated according to the Lausanne atlas (Cammoun et al., 2012), which includes whole-brain parcellations at multiple spatial scales (83, 129, 234, 463, and 1015 regions). Parcellations were defined in native space and co-registered to the first $b = 0$ volume of each subject's diffusion image using a rigid-body transform. To extend gray matter region labels beyond the gray-white boundary, the atlas labels were dilated by 4mm (Gu et al., 2015). Dilation involved filling non-labelled voxels with the statistical mode of neighboring labels. 234 dilated brain regions defined the nodes for each subject's structural brain network, which was represented as a weighted adjacency matrix $A$. Edges were defined where at least one streamline connected a pair of nodes end-to-end. Edge weights were primarily defined by the average FA along streamlines connecting any pair of nodes (Misic et al., 2016; Bohlken et al., 2016). See \textit{Figure 2}.

\subsection*{Functional Module Assignment}
For the 234- and 463-region parcellations, we calculated a \textit{purity index} for each Lausanne label and corresponding voxels in the standard 7-system template image provided by Yeo et al. (2011). This measure quantifies the maximum overlap of cortical Lausanne labels and functional systems defined by Yeo et al. (2011). Each cortical Lausanne label was assigned to a functional system by calculating the non-zero mode of all voxels in each brain region. Subcortical regions were assigned to an eighth, subcortical module. The primary modular partition defined for 234-node networks is shown in \textit{Figure 2}. To determine whether the functionally-defined network partition significantly fit the structural connectivity data beyond chance, we quantified the modularity quality index (formally defined below) of the functional partition imposed on structural brain networks. Briefly, the modularity quality of a network partition quantifies how well that partition maximizes the strength of within-module connections relative to a specified null model. Higher $Q$ values indicate that modules are highly segregated within a network, with strong within-module connectivity and relatively weak between-module connectivity. We performed a permutation test to examine the significance of the modularity quality of the functional partition ($Q_{Yeo}$) imposed on structural connectivity matrices. First, we permuted the assignment of $N$ nodes to functional modules 1000 times, preserving the number of nodes originally assigned to each module. We then calculated the modularity quality $Q_{perm}$ of randomly-defined network partitions imposed on each subject's connectivity matrix, building a null distribution for $Q_{perm}$. We used the calculated mean ($\mu_{Q_{perm}}$) and standard deviation ($\sigma_{Q_{perm}}$) of the null distribution to derive a \textit{z}-score based on the observed $Q_{Yeo}$ for each subject (\textit{z}-score = $\frac{(Q_{Yeo} - \mu_{Q_{perm}})}{\sigma_{Q_{perm}}}$). Finally, we calculated the mean \textit{z}-score across all subjects to assess the significance of $Q_{Yeo}$.

\subsection*{Measures of Modular Segregation}
We calculated the \textit{participation coefficient} to quantify the relative balance of between-module versus within-module connectivity for each brain region. Intuitively, this measure describes the degree to which a brain region integrates information across distinct modules, or the degree to which a brain region shows provincial connectivity among regions in its own module. We define the participation coefficient $P_i$ of node $i$ as
\begin{equation}
P_{i} = 1 - \sum_{m\in{M}} \Big(\frac{k_{i}(m)}{k_{i}} \Big)^2 ,
\end{equation}
where $m$ is a module in a set of modules $M$, and $k_{i}(m)$ is the weight of structural connections between node $i$ and all nodes in module $m$ (Guimera and Amaral 2005; Rubinov and Sporns 2010). Moreover, $P_i$ close to 1 indicates that a brain region is highly integrated with regions in other modules, while a $P_i$ close to 0 indicates that a brain region is highly segregated, with strong connectivity among other regions in its own module. To quantify the segregation of specific modules, we average $P_i$ across all brain regions assigned to the same module. To quantify global network segregation, we average $P_i$ across all nodes in the network. 

\subsubsection*{Alternative Measures of Modular Segregation}

To ensure that our results were not dependent on specific network metrics, we calculated alternative measures of modular segregation. First, we calculated the average strength of all within-module connections (a measure of structural coherence), and the average strength of all between-module connections (a measure of structural integration) in the network (Gu et al., 2015). These metrics provide additional insights into the segregation of information processing within distinct modules, and the degree to which modules are integrated across the network (see \textit{Figure 4}). Alternatively, we calculated the subject-specific modularity quality ($Q$) of group-level functional and structural network partitions. As discussed above, this measure provides an index of how well a network can be decomposed into a hard partition where nodes within the same module demonstrate particularly strong connectivity beyond chance. We also calculated $Q_{subj}$ for subject-specific consensus partitions (see detailed procedure below), which was not dependent on a group-level partition. We calculated the modularity $Q$ of a network partition $S$ based on the following modularity quality function:

\begin{equation} \label{eq:mod}
Q(S)= \frac{1}{2m} \sum_{ij} \Big[A_{ij} - \gamma P_{ij}\Big] \delta(g_{i},g_{j}) ,
\end{equation}
where $m$ is the total weight of $A$, $P$ represents the expected strength of connections according to a specified null model (Newman, 2004), $\gamma$ is a structural resolution parameter that determines the size of modules, and $\delta(g_{i},g_{j})$ is equal to unity when brain regions $i$ and $j$ are assigned to same community $g_{i}$, and is zero otherwise.

\subsubsection*{Community Detection in Structural Brain Networks}

Primary analyses relied on an \textit{a priori} functional partition to define network modules. We additionally estimated network modules directly from the structural connectivity data using community detection procedures. Communities were defined by maximizing the modularity quality function using a generalization of the Louvain heuristic (Blondel et al., 2008; Mucha et al., 2010). Because the Louvain algorithm is degenerate (Good et al., 2010; Sporns and Betzel 2016), it is essential to perform modularity maximization multiple times in order to identify a stable consensus partition that accurately reflects the solutions offered by each optimization. Accordingly, we applied a locally greedy Louvain-like modularity-optimization procedure (Blondel et al., 2008) 100 times for each subject in order to define an ``agreement'' matrix $\mathbf{\mathit{A'}}$ where $A_{ij}'$ was equal to the probability that nodes $i$ and $j$ were assigned to the same community over the 100 iterations. If $\mathbf{\mathit{A'}}$ was deterministic (edge weights were binary), then the algorithm had converged and the resultant partition was defined as the consensus. Otherwise, we performed 100 iterations of modularity optimization on $\mathbf{\mathit{A'}}$ in order to generate a new agreement matrix $\mathbf{\mathit{A''}}$. This procedure was repeated until convergence (Lancichinetti and Fortunato, 2014). When performing modularity optimization on an agreement matrix (e.g., $A'$ or $A''$), we defined an alternative null model $P'$ by permuting community assignments across nodes (Bassett et al., 2013).

Once a consensus partition was identified for each subject, we computed a group-level consensus across the full PNC cohort (n=882). To do this, we used a Louvain-like procedure to detect communities in a group-level agreement matrix $A_{group}'$. Edge weights in $\mathbf{\mathit{A_{group}'}}$ were equal to the proportion of times that each pair of nodes was assigned to the same community across subject-level consensus partitions. As above, 100 iterations of modularity optimization were performed on $\mathbf{\mathit{A_{group}'}}$ until the resulting $\mathbf{\mathit{A_{group}''}}$ became binary, indicating that the algorithm had converged on a group-level consensus partition. Both subject-level and group-level consensus partitions were computed over a wide range of $\gamma$ ($\left[0,4\right]$, in increments of 0.05) to explore variations in community structure. We plotted the number of group-level consensus modules as a function of $\gamma$, and found several plateaus indicating partition stability (Fenn et al., 2009; see \textit{Figure S1}). In order to directly compare the organization of data-driven, modularity-based partitions and the \textit{a priori} functional partition, we quantified the partition similarity using the $z$-score of the Rand coefficient (Traud et al., 2011). For two partitions $X$ and $Y$, we calculated the Rand $z$-score in terms of the total number of node pairs in the network $M$, the number of pairs $M_{X}$ assigned to the same module in partition $X$, the number of pairs $M_{Y}$ that are in the same module in partition $Y$, and the number of pairs of nodes $w_{XY}$ that are assigned to the same module both in partition $X$ and in partition $Y$. The $z$-score of the Rand coefficient is defined by:
\begin{equation}
z_{XY} = \frac{1}{\sigma_{w_{XY}}} w_{XY} - \frac{M_{X}M_{Y}} {M} ,
\end{equation}
where $\sigma_{w_{XY}}$ is the standard deviation of $w_{XY}$. The mean partition similarity is determined by the mean value of $z_{XY}$ over all possible partition pairs for $X \neq Y$. Moreover, $z_{XY}$ denotes the similarity of partitions $X$ and $Y$ beyond chance. \textit{Figure S1} shows the similarity between all group-level structural partitions and the primary functional partition used in this study.

\subsection*{Measures of Network Integration}

For each subject's structural brain network $A$, the topological length or distance of each edge $A_{ij}$ was computed as the reciprocal of the edge weight ($\frac{1}{A_{ij}}$). The path length between any pair of nodes is defined as the sum of the edge lengths along the shortest path connecting them (Rubinov and Sporns, 2010). \textit{Global efficiency} provides a theoretical prediction of how easily information can flow across a network via the shortest path between all pairs of nodes, and is defined by 
 \begin{equation}
E_{glob}(G) = \frac{1}{n} \sum_{i\in{N}} \frac{ \sum_{j\in{N}, j\neq{i}} \bigg(d_{ij}\bigg)^{-1}} {n - 1} ,
\end{equation}
where $n$ is the number of nodes, and $d_{ij}$ is the shortest path length between node $i$ and node $j$. 

To examine the possible role of specific edges as integrative hub connections within the network, we calculated the weighted edge betweenness centrality ($EBC$) for each edge. Edge betweenness identifies important hub connections by providing a measure of how much a given connection participates in the shortest paths of communication through a network, and thus contributes to global efficiency (Brandes, 2001).

 \begin{equation}
 EBC = \sum_{hk} \frac{\rho_{hk}^{ij}} {\rho_{hk}},
 \end{equation}
 
where $\rho_{hk}^{ij}$ denotes the number of shortest paths between nodes $h$ and $k$ that include edge $ij$, and $\rho_{hk}$ denotes the total number of shortest paths between $h$ and $k$. After calculating $EBC$ individually for each weighted network $A_{ij}$ (n=882), we normalized each subjects$'$ $EBC$ values by their maximum observed $EBC$, resulting in a bounded measure $\left[0,1\right]$ (Gong et al., 2009). We calculated the mean normalized $EBC$ for each network edge across subjects, and defined \textit{hub edges} as those connections within the top quartile of normalized edge betweenness across all network edges. Following group-level analysis, which identified a subset of edges that significantly strengthened with age, we performed a permutation-based test to assess whether connections that significantly strengthened with age were enriched for hub edges (see below). 

\subsection*{Group-level analyses}

Prior work has demonstrated that brain development is not a linear process (Paus et al., 1999, Shaw et al., 2006). Accordingly, group-level analyses of structural brain network metrics were flexibly modeled using penalized splines within a General Additive Model (GAM) implemented in the R package ``mgcv'' (https://cran.r-project.org/web/packages/mgcv/index.html; Wood 2004; Wood 2011). Such an approach allows for detection of nonlinearities in the relationship between age and measures of modular segregation without defining a set of functions \textit{a priori} (such as polynomials). Importantly, the GAM estimates nonlinearities using restricted maximum likelihood (REML), and determines a penalty with increasing nonlinearity in order to avoid overfitting the data. Due to this penalty, the GAM only models nonlinearities when they explain additional variance in the data above and beyond linear effects. 
\par

First, we used penalized splines to estimate nonlinear developmental patterns of modular segregation. Within this model we included covariates for sex, head motion, and total network strength. Accordingly, the final model equations for estimating age effects on modular segregation (mean participation coefficient) were as follows:

\vspace{3mm} Modular segregation = spline(age) + sex + motion + total network strength\vspace{3mm}

\noindent An identical model was used when estimating age effects on the participation coefficient of individual brain regions. Similarly, we applied this model across all network edges in order to assess linear and nonlinear age effects on the strength of individual connections. For all analyses, multiple comparisons were controlled using the False Discovery Rate (\textit{q}$<$0.05).

\subsubsection*{Permutation Testing}
We performed permutation-based tests across network edges in order to assess (i) whether the edges that significantly strengthened with age were localized to within-module connections beyond chance, (ii) whether edges that significantly strengthen with age were enriched for hub edges, and (iii) whether these ages had elevated edge betweenness centrality beyond chance. 
\par
First, we permuted a binary edge label specifying whether each edge connects nodes within or between modules 1000 times. Then for permuted samples of within- and between-module edges, we counted the number of edges that were shown to significantly strengthen with age in group-level analysis. We then rank-ordered the number of edges shown to significantly strengthen with age for permuted within-module edge samples, and determined where the observed number of within-module edges that strengthen with age falls relative to this null distribution. 
\par
Second, we evaluated whether edges that significantly strengthen with age were enriched for hub edges. We permuted a binary edge label defining hub or non-hub edges 1000 times. For each permuted sample, we counted the number of edges that significantly strengthened with age in group-level analysis. Then, we rank-ordered the number of permuted hub edges shown to significantly strengthen with age, and compared these values with the observed number of hub edges that strengthened with age. 
\par
Third, we evaluated whether edges that significantly strengthen with age had higher edge betweenness centrality than anticipated by chance. We permuted normalized edge betweenness centrality values 1000 times. For each permuted sample, we calculated the mean $EBC$ of within-module edges and between-module edges that significantly strengthened with age. We rank-ordered the mean $EBC$ of permuted within-module and between-module edges that strengthened with age, and compared these values with the observed means for within- and between-module edges separately (\textit{Figure 6D}).

\subsection*{Methodological Replications}
To verify that observed age-related increases in modular segregation were not simply due to specific network construction choices, we repeated developmental inferences on modular segregation using a variety of other parameters. First, we examined age effects on modular segregation (mean participation coefficient) using a data-driven structural partition identified at the group level (see \textit{Figure S2B.}, \textit{Figure 5B}, and detailed procedure above). Alternatively, we also calculated the modularity quality index for each subject$'$s optimal partition at $\gamma$=2.5 ($Q_{Subj}$), where a higher $Q_{Subj}$ indicates greater modular segegration (\textit{Figure 5C}). Next, we examined modular segregation (mean participation coefficient) using the \textit{a priori} functional partition assigned to a higher-resolution parcellation of the brain (463 nodes instead of 234; see \textit{Figure 5D}). We also measured modular segregation of the functional partition using structural networks with alternative edge weight definitions. While primary analyses focused on FA-weighted structural networks, we also measured modular segregation in streamline-weighted networks (see \textit{Figure 5E}), where edge weights were equal to the number of streamlines connecting a pair of nodes (Bassett et al., 2011), and additionally, where edge weights were defined by streamline density: the number of connecting streamlines divided by the total regional volume of each node pair (Baker et al., 2015; see \textit{Figure 5F}). In addition to examining age-related patterns of modular segregation using alternative network measures and parameters, we also repeated analyses including the following additional covariates in the GAM described above: race, maternal education, handedness, and total brain volume.

\subsubsection*{Relationship Between Modular Segregation and Global Network Efficiency}

First, we examined age-related effects on global efficiency using the same GAM as above:

\vspace{3mm}\noindent Global efficiency = spline(age) + sex + motion + total network strength\vspace{3mm}

\noindent (see \textit{Figure 6A}). The relationship between global efficiency and modular segregation was assessed within a GAM while controlling for age in addition to other covariates described above (\textit{Figure 6B}). Moreover, the model equation was as follows:

\vspace{3mm}\noindent Modular segregation = Global efficiency + spline(age) + sex + motion + total network strength\vspace{3mm}

\noindent To assess whether global efficiency was related to the weight of specific network connections that strengthened with age, we estimated the following GAMs:

\vspace{3mm}\noindent Global efficiency = Average strength of within-module edges + spline(age) + sex + motion + total network strength

\vspace{3mm}\noindent Global efficiency = Average strength of between-module edges + spline(age) + sex + motion + total network strength\vspace{3mm}

\noindent (see \textit{Figure 6E} and \textit{Figure 6F}).

\subsubsection*{Associations with Executive Function}
To examine the association between modular segregation and executive efficiency, we included a spline age term in the model to account for the variance associated with linear and nonlinear age-related changes in executive ability. The final model equation was as follows:

\vspace{3mm}\noindent Modular segregation = spline(age) + executive efficiency + sex + motion + total network strength\vspace{3mm}

\noindent Using the same GAM, we also evaluated the association between the segregation of individual modules (e.g., frontoparietal) and three cognitive efficiency factor scores: executive function, memory, and social cognition (see \textit{Figure 7A}). We note that 2 participants of the full 882 sample had incomplete cognitive datasets: subsequent analyses examining associations between executive function and modular segregation focused on the remaining 880 participants. Visualization of GAM model fits were created using the ``visreg'' package in R (https://cran.r-project.org/web/packages/visreg/). In \textit{Figure 3A}, \textit{Figure 5}, and \textit{Figure 6B}, one outlying datapoint was beyond the axis range, and was excluded for visualization purposes only: group-level analyses and reported results include data points for all subjects.

\subsubsection*{Mediation analyses}

Linear mediation analyses investigated whether age-related improvement in executive function was mediated by modular segregation and/or global efficiency (Preacher and Hayes, 2008). First, we regressed out the effects of nuisance covariates (sex, head motion, and total network strength) on the independent (X), dependent (Y), and mediating (M) variables. The residuals were then used in our mediation analysis. The significance of the indirect effect was evaluated using bootstrapped confidence intervals within the R package ``lavaan'' (https://cran.r-project.org/web/packages/lavaan/). Specifically, we examined the total effect of age on executive performance (c path; \textit{Figure 7B}), the relationship between age and modular segregation (a path), the relationship between modular segregation and executive function (b path), and the direct effect of age on executive efficiency after including modular segregation as a mediator in the model (c$'$ path). The significance of the indirect effect of age on executive function through the proposed mediator (modular segregation) was tested using bootstrapping procedures, which minimize assumptions about the sampling distribution (Preacher and Hayes, 2008). This approach involves calculating unstandardized indirect effects for each of 10,000 bootstrapped samples and calculating the 95$\%$ confidence interval. This procedure was repeated to assess (i) whether the segregation of the frontoparietal module mediated developmental improvements in executive function, (ii) whether the segregation of the default mode module mediated developmental improvements in social cognition, and (iii) whether age-related increases in global efficiency mediated improvements in executive function.

\subsubsection*{Data Visualization}

Network partitions and regional results (\textit{Figure 2}, \textit{Figure 3C}, and \textit{Figure S2}) were visualized on the cortical white matter surface using FreeSurfer visualization tools in MATLAB. While age effects on the participation coefficient for subcortical brain regions are not visualized in \textit{Figure 3C}, these regions were included in all analyses. Brain network visualizations in \textit{Figure 4} and \textit{Figure 6} were generated using BrainNet Viewer (Xia et al. 2013). 

\pagebreak

\section*{Supplemental References}
\begin{enumerate}

\item{Andersson, J.L. and Sotiropoulos, S.N. (2016). An integrated approach to correction for off-resonance effects and subject movement in diffusion MR imaging. Neuroimage, \textit{125},1063-1078.}
\item{Bassett, D.S., Porter, M.A., Wymbs, N.F., Grafton, S.T., Carlson, J.M. and Mucha, P.J. (2013). Robust detection of dynamic community structure in networks. Chaos, \textit{23}, 013142.}
\item{Fenn, D.J. Porter, M.A., McDonald, M., Williams, S., Johnson, N.F., and Jones, N.S. (2009). Dynamic communities in
multichannel data: An application to the foreign exchange market during the 2007-2008 credit crisis. Chaos \textit{19}, 033119.}
\item{Giedd, J.N., Blumenthal, J., Jeffries, N.O., Castellanos, F.X., Liu, H., Zijdenbos, A., Paus, T., Evans, A.C., Rapoport, J.L. (1999). Brain development during childhood and adolescence: a longitudinal MRI study. Nature Neuroscience \textit{2(10)}, 861-3.}
\item{Gong, G., He, Y., Concha, L., Lebel, C., Gross, D.W., Evans, A.C. and Beaulieu, C. (2009). Mapping anatomical connectivity patterns of human cerebral cortex using in vivo diffusion tensor imaging tractography. Cerebral Cortex, \textit{19}, 524-536.}
\item{Good, B. H., Montjoye, Y., and Clauset, A. (2010). Performance of modularity maximization in practical contexts. Physical Review E \textit{81}, 046106.}
\item{Gu, S., Pasqualetti, F., Cieslak, M., Telesford, Q.K., Alfred, B.Y., Kahn, A.E., Medaglia, J.D., Vettel, J.M., Miller, M.B., Grafton, S.T. and Bassett, D.S. (2015). Controllability of structural brain networks. Nature Communications, \textit{6}.}
\item{Lancichinetti, A. and Fortunato, S., 2012. Consensus clustering in complex networks. Scientific reports, \textit{2}.}
\item{Newman, M.E.J., 2004. Analysis of weighted networks. Physical Review E, \textit{70}, 056131.}
\item{Paus, T., Zijdenbos, A., Worsley, K., Collins, D.L., Blumenthal, J., Giedd, J.N., Rapoport, J.L. and Evans, A.C. (1999). Structural maturation of neural pathways in children and adolescents: in vivo study. Science, \textit{283}, 1908-1911.}
\item{Shaw, P., Greenstein, D., Lerch, J., Clasen, L., Lenroot, R., Gogtay, N.E.E.A., Evans, A., Rapoport, J. and Giedd, J. (2006). Intellectual ability and cortical development in children and adolescents. Nature, \textit{440}, 676-679.}
\item{Xia, M., Wang, J. and He, Y. (2013). BrainNet Viewer: a network visualization tool for human brain connectomics. PLoS One, \textit{8}, e68910.}
\end{enumerate}



\end{document}